\documentstyle[12pt]{article}

\begin{document}

\title{
Determination of Exchange Parameters \\
 for CaV$_4$O$_9$
}

\author{ 
Ken'ichi Takano and Kazuhiro Sano$^1$ \\
Toyota Technical Institute, Tenpaku-ku, Nagoya 468 \\
$^1$Department of Physics, Faculty of Education, \\
Mie University, Tsu, Mie 514
}

\date{\hskip 5cm}

\maketitle

\begin{abstract}
      Magnetic susceptibility measured for CaV$_4$O$_9$ is analyzed 
by a method with high temperature expansion.  
      The analysis is consistent with the $g$-value which is close to 2 
and is observed by an ESR experiment.  
      Four exchange parameters in a two-dimensional Heisenberg model 
are determined to represent CaV$_4$O$_9$.  
      All the exchange parameters are about 500 K and strongly 
frustrated. 
      The observed spin gap originates from the frustration. 
\end{abstract}

\noindent
keywords: 
CaV$_4$O$_9$, spin gap, magnetic susceptibility, 
high temperature expansion, two-dimensional antiferromagnet, 
frustration


\section{Introduction}

      Spin gaps in low dimensional materials have been 
investigated by a number of researchers. 
      While origins and effects of spin gaps are fundamentally 
and generally interesting, those for two-dimensional materials 
are especially important in possible relevance to the high 
temperature superconductivity. 
      Recently Taniguchi et al.~\cite{Taniguchi} found a spin gap 
in CaV$_4$O$_9$ by magnetic susceptibility and NMR measurments. 
      CaV$_4$O$_9$ is a layered insulator and has spin degrees of 
freedom at vanadium ions.~\cite{Bouloux} 
      This is the first clear observasion of a spin gap in a 
two-dimensional spin system.  

      Theoretical efforts have been done to describe CaV$_4$O$_9$ 
as a two-dimensional Heisenberg model and to explain the origin of 
the spin
gap.~\cite{Katoh,Ueda,Sano,Troyer,Albrecht,Miyazaki,Gelfand,Fukumoto,Sachdev
,White,Starykh,Albrecht2}  
      Almost models considered are included, as special cases, 
in a Heisenberg model with 4 kinds of exchange parameters;  
$J_e$ for edge-sharing plaquette links, 
$J_e'$ for edge-sharing dimer links, 
$J_c$ for corner-sharing plaquette links and 
$J_c'$ for corner-sharing dimer links as shown in Fig.~\ref{fig:1}.  
      In these studies authors assumed values or constraints 
for the parameters and constructed their theories.  
      Among them, Troyer~\cite{Troyer} performed a quantum 
Monte Carlo simulation and argued that the system does not form 
a spin gap in the unfrustrated homogenous case of $J_e = J_e' \ne 0$ 
and $J_c = J_c' = 0$.  
      The spin gap is possibly formed by inhomogeneity:  
i. e. the difference between $J_e$ and $J_e'$.  
      It is also possible that the spin gap originates from frustration 
due to nonzero $J_c$ and $J_c'$.  
%
\begin{figure}
\vskip 6cm
\caption{
      Lattice structure for vanadium spins in a layer of 
CaV$_4$O$_9$.  
      The Heisenberg model examined in text includes 
4 dependent exchange parameters $J_e$ (bold solid line), 
$J_e'$ (solid line), $J_c$ (bold dashed line) and 
$J_c'$ (dashed line).}
\label{fig:1}
\end{figure}

      In a previous paper,~\cite{Sano} we roughly estimated $J_e$ and 
$J_c$ on the condition of $J_e' = J_e$ and $J_c' = J_c$. 
      We fitted a high-temperature expansion to the order of $1/T^3$ 
for the susceptibility to experimental data.  
      The result was $J_e \simeq$ 610 K and $J_c \simeq$ 
150 K with keeping the opposite possibility.~\cite{Sano2}  
      Calculation of numerical diagonalization shows that the 
frustration among interactions with $J_e$ and $J_c$ enhances 
the spin gap in comparison with the case of $J_c =0$.  
      This enhancement agrees with the result of a perturbation 
calculation.~\cite{Ueda} 

      Most recent experiments give more detailed information on this 
material: 
      (i) Neutron scattering measurement~\cite{Kodama} suggests 
$J_c \displaystyle{ \mathop{>}_{\sim}} J_e$.  
      (ii) ESR measurement~\cite{Kobayashi} precisely determined 
the $g$-value as $g$ = 1.96, which is rather close to 2.  
      The Curie constant is then $C$ = 0.003713 emu/g. 
      As for our previous paper, experiment (i) rather supports 
the opposite possibility for the values of the exchange parameters; 
i. e. $J_c \simeq$ 610 K and $J_e \simeq$ 150 K.  
     On the other hand, we have used $g$ = 2 of a free spin for 
$g$-value without explicit notice there.  
      This $g$-value is close to the truth.  
      Gelfand et al. estimated the exchange parameters as 
$J_e \simeq J_e' \simeq$ 190 K and $J_c \simeq J_c' \simeq J_e/2$ 
by various expansions.  In the estimation, they determined the $g$-value
as $g$ = 1.77.  
      The difference between this value and the experimental value seems 
to be hardly explained.  

      In view of the above, we expect that the previous estimation 
for $J_c$ and $J_e$ is roughly realistic, if we exchange the values 
of them.  
      However, the estimation still includes unsatisfactory points.  
      First we examined a Heisenberg model including only two 
independent exchange parameters, $J_e$ and $J_c$.  
      It is at least necessary to independently treat parameters 
for plaquette and dimer links in order to determine the origin 
of the spin gap.  
      Second we have a room to improve the fitting method.  
      We obtained an expansion of the susceptibility of the 
Heisenberg model up to a finite order and fitted it directly 
to the experimental data.  
      This method is rather usual for high temperature fitting.  
      However the estimation is rough unless fairly high order 
expansion is carried out, and it is hard to calculate high order 
terms of expansion for a system with many independent 
exchange parameters.
      
      In this paper, we  determine the exchange parameters of 
the Heisenberg model to describe CaV$_4$O$_9$ in a novel 
fitting method. 
      Following this method, we first construct an experimental 
formula for the magnetic susceptibility in power series of 1/T. 
      This procedure is explained in detail in section 2. 
      On the other hand, we obtain a high temperature expansion of 
the susceptibility for a Heisenberg model with 4 independent 
exchange parameters;  the expansion coefficients are functions of 
the exchange parameters.  
     We determine the parameters by fitting the expansion 
coefficients to the coefficients of the experimental formula.  
      The fitting is carried out in section 3. 
      The result shows that the origin of the spin gap for this 
material is the frustration rather than the inhomogeneity.  
      Discussion is devoted in section 4. 


\section{Experimental formula for susceptibility 
at high temperatures}

      The experimentally obtained magnetic susceptibility, 
$\chi^{EXP}(T)$, is a set of data for 
$T \displaystyle{\mathop{<}_{\sim}}$700 K.  
      We construct an experimental formula for the data 
in a power series of $1/T$.  
      The formula will be used to compare the experiment 
to a theory at high tempratures.  
      To make expansion coefficients dimensionless we use 
$x=T_0/T$ as the expansion parameter, where $T_0$ is 
an arbitrary constant with the dimension of temperature.  
      We hereafter employ $T_0$ = 700 K without spoiling 
generality.  
      The formula is then written as 
\begin{eqnarray} 
\label{chi} 
     \chi(T) &=& \lim_{n \rightarrow \infty} \chi^{(n)}(T) , \nonumber \\
     \chi^{(n)}(T) &=& \frac{C}{T} 
                                [ 1 + \sum_{m=1}^{n} A_m x^m ] 
\end{eqnarray}
with Curie constant $C$ deteremined by the ESR measurement.
      Here the expansion coefficients $A_m$'s are fitting 
parameters.~\cite{Fitting}  

      To determine $A_m$'s we introduce a novel fitting method as 
will be explained below. 
      We first define the quantity 
\begin{equation}
\label{phiFITd}
   \phi_m(x) = A_m + \sum_{l=1}^{\infty} A_{m+l} x^l 
\end{equation}
for each $m$.  
      It is important that the quantity reduces to coefficient 
$A_m$ in the high temperature limit:  
\begin{equation}
\label{phiFITh}
      \phi_m(0) = A_m.  
\end{equation}
     $\phi_m(x)$ is also defined by the recursion equation: 
\begin{eqnarray}
\label{phiFIT}
   \phi_0(x) &=& \frac{T}{C} \ \chi(T) , \nonumber \\
   \phi_m(x) &=& (\phi_{m-1}(x) - A_{m-1}) 
   \frac{1}{x}
\end{eqnarray}
for $m=0, 1, 2, \cdots$ with $A_{0}$ = 1.  
      The experimental data corresponding to $\phi_m(x)$ 
is similarly defined in the recursive transformation:  
\begin{eqnarray}
\label{phiEXP}
   \phi^{EXP}_0(x) &=& \frac{T}{C} \ \chi^{EXP}(T) , \nonumber \\
   \phi^{EXP}_m(x) &=& (\phi^{EXP}_{m-1}(x) - A_{m-1}) 
   \frac{1}{x} .
\end{eqnarray}
      Here $\phi_m$ fits $\phi^{EXP}_m$ because 
$\chi$ is constructed to fit $\chi^{EXP}$.  
      It is noticed that $\phi_m$ is only formally introduced 
since the starting function $\chi$ is now unknown and 
is the object which we will obtain finally. 
      On the other hand we have the data $\chi^{EXP}$ and so 
$\phi^{EXP}_m$ is actually obtained as is mentioned below.  

      Now we determine $\{A_m\}$ together with 
$\{\phi^{EXP}_m(x)\}$ one by one.  
      For $m$ = 0, we have $\phi^{EXP}_0(x)$ by multiplying 
$T/C$ to $\chi^{EXP}(T)$ and have $A_0$ = 1 by the definition.  
      Next we obtain $\phi^{EXP}_1(x)$ by the transformation 
$\phi^{EXP}_1(x) = (\phi^{EXP}_0(x) - A_0) /x$ 
in eq.~(\ref{phiEXP}) with $m$ = 1.  
      To obtain $A_1$ we use a simple fitting function 
$f_1(x) = a_1 + b_1 \exp(- c_1 x)$ and determine 
the parameters $a_1$, $b_1$ and $c_1$ to make $f_1(x)$ 
fit $\phi^{EXP}_1(x)$ by the least square method.  
      Using these parameters, $A_1$ is given by 
$A_{1}$ = $f_1(0)$ = $a_1 + b_1$ corresponding to 
$A_{1} = \phi_1(0)$ in eq.~(\ref{phiFITh}) with $m$ = 1.  
      By repeating this process, we obtain $\phi_m(x)$ 
and $A_m$ for arbitrary $m$:  
      When $\phi_{m-1}(x)$ and $A_{m-1}$ have been known, 
$\phi^{EXP}_m(x)$ is given by eq.~(\ref{phiEXP}).  
      Then we make function 
\begin{equation}
\label{fm}
   f_m(x) = a_m + b_m \exp(- c_m x) 
\end{equation}
fit $\phi^{EXP}_m(x)$ and determine parameters 
$a_m$, $b_m$ and $c_m$.  
      $A_m$ is obtained by $A_m = f_m(0) = a_m + b_m$ 
corresponding to eq.~(\ref{phiFITh}).  
      Thus we can determine any coefficient $A_m$ inductively.  

      In Fig.~\ref{fig:2}, we show $\phi^{EXP}_m(x)$ along with 
$f_m(x)$ for $m$ = 1 to 4.
      Here we have used a weight function $\exp(- 1000/T)$ 
for fitting by the least square method.  
      Optimal values for $a_m$, $b_m$, $c_m$ and $A_m$ 
are shown in Table~\ref{table:1}.  
      In principle we can obtain coefficient $A_m$ for any $m$.  
      However, experimental data $\phi^{EXP}_m(x)$ becomes 
dispersive as $m$ increases, so that $A_m$ with very large $m$ 
cannot be obtained.  
      When we change the weight function, the values of $A_m$'s change 
with a strong correlation:  
      $|A_m|$'s have a tendency to increase or decrease simultaneously. 
      The error depending on the choice of the weight function seems 
to be roughly several percents.  
\begin{figure}
\vskip 8cm
\caption{Transformed experimental data $\phi^{EXP}_m(x)$ and 
fitting function $f_m(x)$ for $m$ = 1 to 4.} 
\label{fig:2}
\end{figure}
\begin{table} 
\begin{center} 
\caption{Fitting parameters and coefficients $A_m$'s.} 
\label{table:1} 
\begin{tabular}{@{\hspace{\tabcolsep}\extracolsep{\fill}}ccccc} \hline
$m$   & $a_m$  & $b_m$  & $c_m$  & $A_m$ \\ \hline
1  & -0.20907 	& -0.76379  & 1.2496   & -0.97286 \\
2  &  0.15067 	&  0.77787  & 0.68028  &  0.92854 \\
3  & -0.10418 	& -0.42303  & 0.41366  & -0.52721 \\
4  &  0.04797	&  0.12828  & 0.29549  &  0.17624 \\
5  & -0.01365	& -0.02554  & 0.26649  & -0.03918 \\
6  &  0.00249	&  0.00484  & 0.26474  &  0.00733 \\
7  & -0.00024	& -0.00104  & 0.15695  & -0.00127 \\
8  &  0.00002	&  0.00014  & 0.08627  &  0.00016 \\ \hline
\end{tabular}
\end{center}
\end{table}

      We have obtained values for $A_m$'s within some accuracy.  
      They are independent of $n$ if $n$ is sufficiently large.  
      It is now instructive to substitute the values for $A_m$'s 
into $\chi^{(n)}$ in eq.~(\ref{chi}) with small $n$ as well as 
large $n$.  
      Results for several values of $n$ are shown in Fig.~\ref{fig:3}.  
      $\chi^{(n)}$ approaches to $\chi^{EXP}$ as $n$ increases, 
confirming the validity of this method.  
\begin{figure}
\vskip 6cm
\caption{ 
$\chi^{(n)}$'s as functions of $T$. 
They are given by eq.~(\ref{chi}) with $A_m$'s in Table \ref{table:1}.
Data of the experimental susceptibility $\chi^{EXP}$ are also 
shown by open circles. 
}
\label{fig:3}
\end{figure}


\section{High temperature expansion for a Heisenberg model 
and determination of exchange parameters} 

      We assume that spins at vanadium sites in a layer of 
CaV$_4$O$_9$ are described by a two-dimensional Heisenberg 
model which is represented in Fig.~\ref{fig:1}. 
      The Hamiltonian is written as
\begin{equation}
\label{Ham}
   H = \sum_{<i,j>} J_{ij} {\bf S_{i}} \cdot {\bf S_{j}} ,
\end{equation}
where ${\bf S_{i}}$ is the spin at site $i$.
      The exchange parameter $J_{ij}$ is $J_e$, $J_e'$, $J_c$ 
or $J_c'$ if it corresponds to a link indicated in Fig.~\ref{fig:1} 
and is zero otherwise.  

      By the high temperature expansion, magnetic susceptibility 
of a Heisenberg model is generally written in the following form:  
\begin{equation}
\label{chiHTE}
   \chi^{HTE}(T) = \frac{C}{T} 
            [ 1 + \sum_{m=1}^{\infty} F_m x^m ] 
\end{equation}
with $x = T_0/T$ and the experimentally determined Curie constant $C$. 
      We have used $T_0$ = 700 K as in the previous section.  
      The coefficients $F_m$'s are functions of the exchange parameters 
and are calculated by the standard diagramatic method.~\cite{Rushbrooke} 

      In the present model of eq.(\ref{Ham}), the coefficients $F_m$'s are 
functions of $J_e$, $J_e'$, $J_c$ and $J_c'$.  
      We obtained them for $m$ = 1, 2 and 3 as  
%
\begin{eqnarray}
\label{Fm-J}
   F_1 T_0    &=& - \frac{1}{4} (2J_e + J_e' + 2J_c + J_c') ,  \nonumber \\
   F_2 T_0^2 &=& \frac{1}{4^2} [(4J_e - J_e') J_e' 
           + 2 (2J_e + J_e') (2J_c + J_c') + (4J_c - J_c') J_c'] ,  \nonumber \\
   F_3 T_0^3 &=& \frac{1}{3 \cdot 4^3} 
           [(8J_e^3 - 12J_e^2J_e' + 6J_eJ_e'^2 + J_e'^3) 
         + 3J_c' (6J_e^2 - 12J_eJ_e' + J_e'^2)  \nonumber \\
        &-& 6J_c (4J_e^2 + 7J_eJ_e' - J_e'^2)  
         - 3 (4J_c^2 + 12J_cJ_c' - J_c'^2) (2J_e + J_e')  \nonumber \\
         &+& (8J_c^3 - 12J_c^2J_c' + 6J_cJ_c'^2 + J_c'^3) ] .
\end{eqnarray}
%

      If the Hamiltonian (\ref{Ham}) completely describes the material, 
we have $\chi^{HTE}(T) = \chi(T)$ or $F_m = A_m$ for all $m$, 
where $A_m$'s are in Table~\ref{table:1}. 
      However the Hamiltonian may only approximately describe 
the real material.  
      We can estimate optimal values for the exchange parameters 
by minimizing the $k$th deviation 
\begin{equation}
\label{D}
            D_k = \sqrt{\frac{1}{k} \sum_{m=1}^k (F_m - A_m)^2} . 
\end{equation}

      In the present case, the exchange parameters are $J_e$, $J_e'$, $J_c$ 
and $J_c'$. 
      We numerically minimized $D_3$ with $F_m$'s in eq.(\ref{Fm-J}) and 
obtained the optimal values as 
\begin{eqnarray}
\label{Je-Jc} 
       J_e \simeq 480 \ {\rm K} , \ J_e' \simeq 530 \ {\rm K} , \nonumber \\
       J_c \simeq 580 \ {\rm K} , \ J_c' \simeq 540 \ {\rm K} . 
\end{eqnarray}
      This estimation is the main result of this paper. 
      The result shows that inhomogeneity is weak and frustration is 
strong in this material.  
      Thus the observed spin gap originates from frustration. 
      These values yield deviation $D_3 \simeq$ 0.11 and possible 
reasons of the deviation are discussed in the next section.


\section{Discussion} 

      We introduced a novel method to estimate exchange parameters. 
      This method is fairly general and so applicable to various systems 
when we compare a Hamiltonian to experiment data.  
      In the first step of this method, we constructed an experimental 
formula for magnetic suscestibility in the power series of $T_0/T$. 
      To examine the accuracy of this method, we applied it to a known 
case: a Heisenberg model on a simple square lattice. 
      The formula of high temperature expansion for this model is already 
calculated.~\cite{Rushbrooke}. 
      Using this formula instead of experimental data, we obtained 
coefficients $A_m$'s in the way of section 2. 
      We confirmed that the coefficients approximately repropduce the 
original formula. 

      We applied the method to CaV$_4$O$_9$ to describe it by 
a two-dimensional Heisenberg model (\ref{Ham}). 
      The optimal values for exchange parameters are shown 
in eq. (\ref{Je-Jc}). 
      The result shows that the strong frustration opens a spin gap 
in this material. 
      The optimal values yield a deviation of $D_3 \simeq$ 0.11. 
      We cannot decide now whether or not this deviation is only 
an error within this method itself. 
      We point out other possible origines of the deviation. 
      It is possible that the deviation comes from unexpected components, 
e. g. VO$_2$, CaV$_2$O$_5$, CaV$_3$O$_7$, included in samples of 
CaV$_4$O$_9$. 
      The deviation also possibly means that the material includes some 
degree of freedom which cannot be described by a Heisenberg model 
(\ref{Ham}); 
      it might come from itinerant effect or effect of degeneracy of atomic 
orbitals.  
      Anyway what we have done is to approximately represent CaV$_4$O$_9$ 
by the Heisenberg model (\ref{Ham}). 

      The values of exchange parameters are plausible if similar materials 
have similar values. 
      In CaV$_2$O$_5$ we estimated $J \sim$ 600 K by fitting the formula 
for the one-dimensional Heisenberg model to experiment. 
      This value is similar to a typical exchange parameter $\sim$ 500 K 
for CaV$_4$O$_9$. 
      In CaV$_3$O$_7$, $J_c \approx J_e$ is argued~\cite{Harashina} 
by using a theoretical result.~\cite{Kontani} 
      These results are actually similar to our result for CaV$_4$O$_9$. 

      Using the values of the exchange parameters, we calculated the 
spin gap of the Hamiltonian (\ref{Ham}) by the numerical diagonalization. 
     The extrapolated spin gap $\Delta$ is given as $\Delta \sim$ 170 K. 
     Considering the 10 \% deviation of the coefficients and errors for 
extrapolation, this result seems to be consistent with the observed 
value of $\Delta \sim$ 110 K.


\section*{Acknowledgements}
      We thank Masatoshi Sato, Satoshi Taniguchi, Hiroshi Harashina 
and Katsuaki Kodama for useful discussion and presentation of 
detailed experimental data before publication.  
      We also thank K. Kubo for useful discussion on high 
temperature expansion of Heisenberg models.  
      One of us (K. T.) carried out this work partially in Department 
of Physics of Nagoya University as a Guest Associate Professor.  
      This work is partially supported by the Grant-in-Aid for 
Scientific Research from the Ministry of Education, Science 
and Culture, Japan.  
      The computaion in this work was carried out partially by 
using facilities of the Supercomputer Center, Institute 
for Solid State Physics, University of Tokyo.  



\begin{thebibliography}{99}
\bibitem{Sano2}
      The first and the second order terms in high-temperature 
expansion formula of the susceptibility are symmetric with 
respect to the exchange between $J_e$ and $J_c$.
      Hence the opposite possibility of $J_e$ = 150 K and 
$J_c$ = 610 K remains.
\bibitem{Fitting}
      One might expect that $A_m$'s are determined by fitting 
$\chi^{(n)}$ with finite $n$ to $\chi^{EXP}$ directly by the least 
square method.  
      However it fails even if the $n$ is fairly large.  
      Fitting $\chi^{(n)}$ to $\chi^{EXP}$ gives $A_m$'s which are 
strongly depend on $n$ and does not seem to converge. 
      It is due to the lack of data in the high temperature regime 
($0 < x \displaystyle{ \mathop{<}_{\sim}} 1$) in $\chi^{EXP}$. 
      The shape of polynomial $\chi^{(n)}$ in this regime sometimes 
oscillates unrealistically. 
      We have introduced another fitting method to overcome this 
difficulty.
\end{thebibliography}
\end{document}